# Spontaneous rotational symmetry breaking induced by electronic instability in the normal state of La$_{1-x}$Sr$_x$NiO$_2$

Qiang Zhao,[1,*] Rui Liu,[1,*] Wen-Long Yang,[1] Xue-Yan Wang,[1] Jia-Kun Luo,[1] Jing-Yuan Ma,[1] Fang-Hui Zhu,[1] Cheng-Xue Chen,[1] Mei-Ling Yan,[1] Rui-Fen Dou,[1] Chang-Min Xiong,[1] Chi Xu,[1] Xing-Ye Lu,[1] Hai-Wen Liu,[1,2] Ji-Kun Chen,[3] Zhi-Ping Yin,[1,2†] and Jia-Cai Nie [1,2†]

[1]School of Physics and Astronomy, Beijing Normal University, Beijing 100875, People's Republic of China

[2]Key Laboratory of Multiscale Spin Physics, Ministry of Education, Beijing Normal University, Beijing 100875, People's Republic of China

[3]School of Materials Science and Engineering, University of Science and Technology Beijing, Beijing 100083, China

*These authors contributed equally to this work.

† Contact authors: jcnie@bnu.edu.cn; yinzhiping@bnu.edu.cn

**The spontaneous rotational symmetry breaking (RSB), a hallmark phenomenon in cuprate and iron-based high-temperature superconductors, is believed to intimately connected to superconductivity, both of which originate from interactions among different degrees of freedoms and competing quantum states. Understanding RSB is pivotal for unraveling the microscopic origin of unconventional superconductivity. Although infinite-layer nickelates (ILNs) share similar crystalline structure and the same nominal 3$d$-electron configurations with cuprates, they have significant differences in Fermi surface topology, electronic band characteristics, and charge order. These distinctions make ILNs an ideal platform for studying RSB in unconventional superconductors. Through angular-resolved resistivity measurements within a large temperature and doping range, we identify pronounced RSB signatures near doping concentrations $x$=0.05 and 0.25. Based on the strongly correlated electronic structures from combined density functional theory and dynamical mean field theory calculations, we find that the calculated electronic susceptibility has a peak structure at the corresponding doping concentration, indicating pronounced electronic instabilities which drive RSB. Detailed analysis of the electronic susceptibility demonstrates that the van Hove singularity at the Fermi level significantly contributes to the electronic instability at 0.05 Sr doping. Our findings reveal the important role of electronic correlation, Van Hove singularity, and Fermi surface nesting in the emergence of RSB. Our work not only deepens the understanding of electronic behavior in ILNs, but also provides new ideas and methods for exploring RSB in other unconventional superconductors.**

The spontaneous rotational symmetry breaking (RSB) in electronic fluids, characterized by a diminished symmetry relative to the host lattice, stands as a pivotal research frontier in the realm of unconventional superconductivity[1-3]. This phenomenon manifests ubiquitously across diverse material platforms, including unconventional superconductors, such as cuprate[4-6], iron-based[7-9], kagome[10,11] and two-dimensional (2D) systems[12-15]. The underlying mechanism of the RSB in unconventional superconductors is complex and multifaceted, frequently linked to unidirectional charge density waves (CDWs)[16], stripe phases[17], nematic order[18], electronic nematicity[4], and interplay between the superconductivity and other emerging quantum states[19].

The uncovering of infinite-layer nickelate (ILN) superconductors, as an analog of cuprate superconductors, has rapidly attracted wide spread attention[20-34]. Although ILNs share similar crystalline structure and $3d$-electron configuration with cuprates, emerging evidence reveals critical distinctions in their long-range antiferromagnetic order[35-38], self-doping as well as electronic structure and Fermi surface topology[39-41]. It is particularly controversial whether there is a CDW in parent and underdoped ILN systems, where conflicting reports exist regarding their existence and stability[24,42-47].

RSB has been probed through diverse experimental approaches: anisotropic transport measurements[4,6,48-50], Nernst effect[51], scanning tunneling spectroscopy[52], and angle-resolved photoemission spectroscopy[11]. Among these, anisotropic resistivity measurements offer unique advantages in sensitivity and experimental accessibility. Pioneering work by Wu *et al.*[5] developed angle-resolved resistivity (ARR) with enhanced precision over traditional Montgomery methods[53], revealing robust fourfold symmetry breaking across wide temperature and doping ranges in cuprates, suggesting electronic nematicity as the origin of RSB. Intriguingly, contrasting behavior emerges in ILNs: Ji *et al.*[19] observed isotropic superconducting states, through the Corbino-disk configuration, in $Nd_{0.8}Sr_{0.2}NiO_2$ films at low fields, with sequential emergence of fourfold ($d$-wave pairing) and twofold (charge stripes) symmetries under increasing magnetic fields. This dichotomy between ILNs and cuprates motivates systematic ARR investigations of RSB across different electronic states.

In strongly correlated electron systems, RSB typically results from the intricate interactions among different degrees of freedom, such as spin, charge, orbital, and lattice[54-61]. As far as we know, theoretical studies have been conducted on the instabilities of spin [62,63], charge[46,63], and lattice[46,63] in ILNs. These theoretical studies suggest that there are strong instabilities in the parent compounds. However, substantial discrepancies emerge between experimental observations and theoretical predictions. Parzyck *et al.*[24], and Raji *et al.* [47] employed a combination of scanning transmission electron microscopy and resonant X-ray scattering techniques in their experiments. They found that there were periodic excess apical oxygens in the parent ILNs which form a triple-periodic superlattice and can explain the charge order features observed by previous experiments. Undoubtedly, this experimental finding poses a challenge to the current theoretical work on ILNs.

In this work, we employ pulsed laser deposition (PLD) to synthesize $La_{1-x}Sr_xNiO_2$ ($x$=0, 0.05 0.1, 0.15, 0.2, 0.25, 0.27, 0.3) epitaxial thin films on $SrTiO_3$(001) substrates. In order to avoid excess apical oxygens in ILNs, we do not grow any capping layers on the nickelate films, and we extend the reduction time of $La_{1-x}Sr_xNiO_3$ by $CaH_2$ powder. Detailed synthesis protocols, lithographic processes, and ARR measurement methodologies are presented in the Methods and Extended Data. To isolate intrinsic electronic anisotropy, we implement a rigorous control experiment using Nb films, confirming that extrinsic factors like thickness variation and lithographic artifacts induce negligible anisotropy. Strikingly, underdoped $La_{0.95}Sr_{0.05}NiO_2$ and overdoped $La_{0.75}Sr_{0.25}NiO_2$ exhibits pronounced RSB strength. This contrasts sharply with the isotropic response of Nb, enabling unambiguous identification of intrinsic RSB in nickelates. The calculated electronic susceptibility from the correlated electronic structures has the same trend as the experimental RSB strength over the studied doping range, which suggests electronic correlation, van Hove singularity (VHS), and Fermi surface nesting (FSN) are important for the formation of RSB. Through comparative analysis of parent-to-overdoped $La_{1-x}Sr_xNiO_2$ films, we established the evolution of electronic instability across the phase diagram, providing critical insights into the RSB of unconventional superconductors.

## Results

### The crystalline and band structures of $La_{1-x}Sr_xNiO_2$

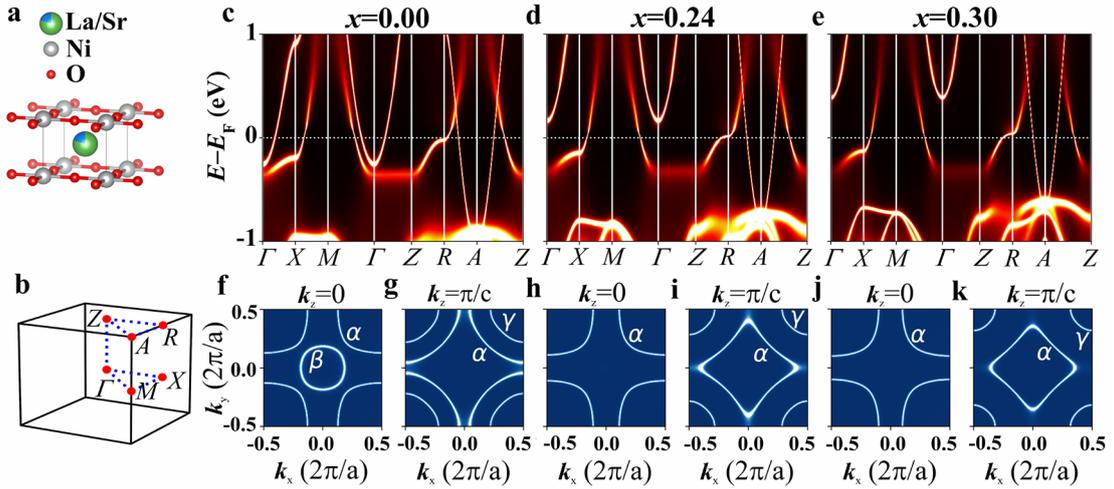

**Fig. 1 | Crystalline and band structures.** (**a**) Crystal structure and (**b**) three-dimensional Brillouin zone of the ILN $La_{1-x}Sr_xNiO_2$. (**c-k**) The DFT+DMFT band structures (**c-e**) and two-dimensional Fermi surfaces (**f-k**) in the $\Gamma$ plane ($k_z=0$) and $Z$ plane ($k_z=\pi/c$) of ILNs with different hole doping concentrations $x$=0, 0.24, 0.3 (from left to right, respectively).

As shown in Fig. 1a, $La_{1-x}Sr_xNiO_2$ has a crystal structure similar to $CaCuO_2$, presenting an infinite-layer structure, with a space group *P4/mmm*. The three-dimensional Brillouin zone of $La_{1-x}Sr_xNiO_2$ is shown in Fig.1b, where the red dots represent high symmetry momentum. We calculate the band structures and Fermi surfaces of $La_{1-x}Sr_xNiO_2$ from $x$=0 to $x$=0.3 using density functional theory combined with dynamical mean field theory (DFT+DMFT) at the temperature 116 K, as shown in Figs. 1c-k. In $LaNiO_2$, there are three bands crossing the Fermi level, where two La-5$d$ orbitals form electron pockets $\beta$ and $\gamma$ near the $\Gamma$ point and $A$ point, respectively, and the Ni-3$d_{x^2-y^2}$ orbital forms a hole pocket $\alpha$ near the $M$ point. As the doping concentration increases, the size of the $\beta$ pocket rapidly decreases, while the $\gamma$ electron pocket changes slowly. The α hole pocket gradually enlarges with the increase in doping concentration. The evolution of the electronic structure with doping obtained from our calculations is consistent with angle-resolved photoemission spectroscopy measurements[64] of $La_{0.8}Sr_{0.2}NiO_2$ as well as previous theoretical calculations[28,46,62-66].

**Rotational symmetry breaking in the infinite-layer nickelates**

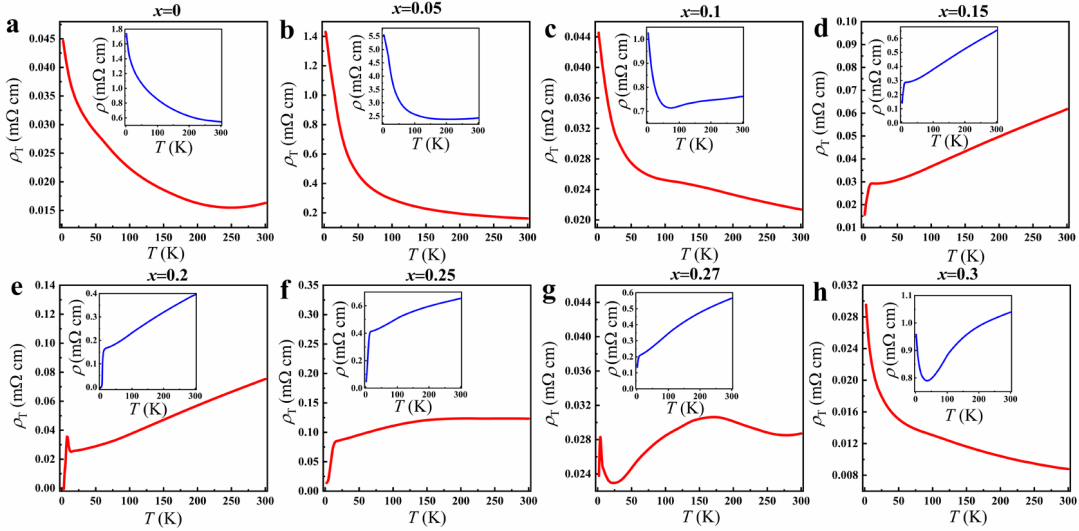

**Fig. 2 | Temperature dependence of the longitudinal and transverse resistivity.** The temperature dependence of the transverse resistivity $\rho_T(T)$ of $La_{1-x}Sr_xNiO_2$ thin films for **(a)** $x=0$, **(b)** $x=0.05$, **(c)** $x=0.1$, **(d)** $x=0.15$, **(e)** $x=0.2$, **(f)** $x=0.25$, **(g)** $x=0.27$, **(h)** $x=0.3$, respectively. The inset shows the corresponding longitudinal resistivity $\rho(T)$.

Figs. 2a-h depict the transverse resistivity $\rho_T(T)$ as a function of temperature, and the insets illustrate the temperature-dependence of the longitudinal resistivity $\rho(T)$. Referring to Figs. 2e and 2g ($x=0.2$ and $0.27$), the $\rho_T(T)$ curve shows a minor peak, which is significantly less pronounced than those observed in cuprate superconductors[4,5], in the vicinity of the superconducting transition temperature. This suggests that the superconducting anisotropy in ILNs is significantly weaker than that in cuprate superconductors. Understanding the origin of the superconducting anisotropy at these doping concentrations requires further experiments at additional doping concentrations[19,67]. While $\rho(T)$ exhibits superconductivity at $x=0.15$ and $0.25$, no peak is observed in $\rho_T(T)$ near the superconducting temperature, suggesting the absence of anisotropy in the superconducting state at these doping concentrations. Moreover, the values of $\rho_T(T)$ at $x=0.05$ and $0.25$ are significantly higher than those at other doping concentrations. This is strong evidence that the normal state in the vicinity of doping concentrations $x=0.05$ and $0.25$ exhibits considerable RSB strength $\eta = \Delta\rho/\bar{\rho}$ (see Methods).

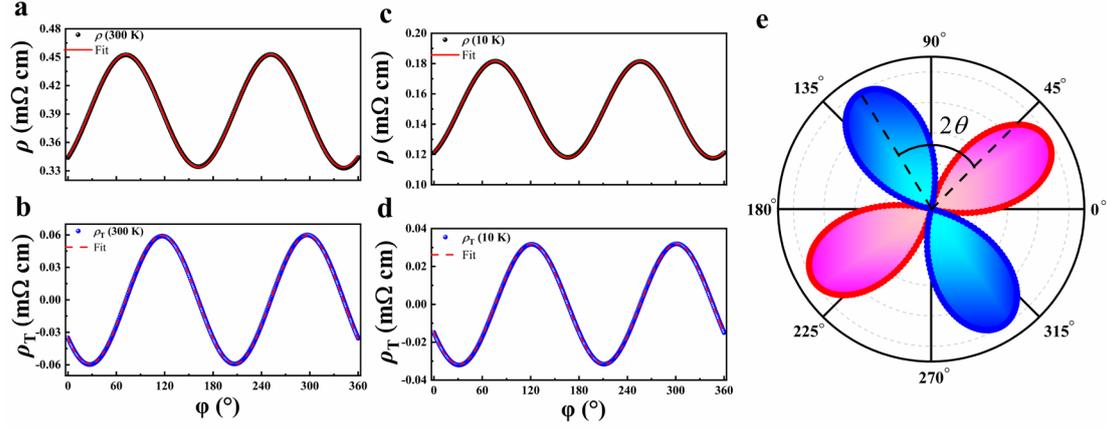

**Fig. 3 | Angle, temperature and doping dependence of the longitudinal and transverse resistivity.**
(a-d) The angle dependence of the **longitudinal resistivity** $\rho$ (**a**, **c**) and **transverse resistivity** $\rho_T$ (**b**, **d**) in $La_{0.8}Sr_{0.2}NiO_2$ at 300 K (**a**, **b**) and 10 K (**c**, **d**), respectively. The experimental data of $\rho$ and $\rho_T$ is represented by black and blue dots, respectively. (**e**)The angular dependence of transverse resistivity $\rho_T$ at 10 K in the polar coordinate system. Notably, both $\rho(\varphi)$ and $\rho_T(\varphi)$ exhibit significant angular oscillations at room temperature and extremely low temperature, with $T = 300$ K (**a**) and 10 K (**b**). The dashed red curves are well fitted by equation $\rho_T(\varphi) = \Delta\rho\sin[2(\varphi-\theta)]$ with $\Delta\rho$=59.44 μΩ cm and 31.82 μΩ cm, $2\theta$ =71.76° and 76.15°, respectively. The solid red curves are well fitted by equation $\rho(\varphi) = \bar{\rho}+ \Delta\rho\cos[2(\varphi-\theta)]$ with $\bar{\rho}$=393.09 μΩ cm and 149.62 μΩ cm, $\Delta\rho$=59.53 μΩ cm and 31.84 μΩ cm, $2\theta$=71.67° and 76.26°, respectively. In the polar coordinate system, the $\rho_T$ of $La_{0.8}Sr_{0.2}NiO_2$ varies in a petal shape with angle, similar to the symmetry of $d$-electron orbitals. The radial distance signifies the magnitude of $\rho_T$, where positive values are represented in blue and negative values in pink. Phase offset $2\theta$ is the angle between the shown extremum and 45°, which is the angle between the two black dashed lines.

If the in-plane crystal exhibits $C_4$ symmetry, $\rho(\varphi)$ must be isotropic, and the zero-field $\rho_T(\varphi)$ should be zero at all angles. Conversely, if the symmetry detected through electrical transport measurements diminishes to $C_2$, $\rho_T(\varphi)$ must obey $\rho_T(\varphi) = \Delta\rho\sin[2(\varphi-\theta)]$ ($2\theta$ is phase offset as shown in Fig. 3e). For anisotropic in-plane transport, the formula of the resistivity tensor is detailed in Methods. Evidently, from Figs. 3a and 3b, both $\rho(\varphi)$ and $\rho_T(\varphi)$ oscillate with a 180° period at both room temperature and extremely low temperatures, exhibiting the same angular period and amplitude. As anticipated, there is a phase shift of $\Delta\varphi$=45° between $\rho(\varphi)$ and $\rho_T(\varphi)$, and the oscillation of $\rho(\varphi)$ is superimposed on the background of the average resistivity $\bar{\rho}$. Experimental data presented in the Supplementary Information III demonstrates that in the normal state, samples with different doping levels exhibit orthogonal anisotropy. However, the same methodology employed to detect metal Nb in the Supplementary Information II failed to reveal any anisotropy, thereby confirming that this property is intrinsic. A representative $\rho_T(\varphi)$ of $La_{0.8}Sr_{0.2}NiO_2$ in polar coordinates at 10 K is presented in Fig. 3c. The petal-like pattern, similar to the $d$-wave superconducting order parameter symmetry observed in cuprate superconductors, serving as further evidence for the RSB of ILNs into $C_2$ symmetry.

# Van Hove singularity in the La$_{0.95}$Sr$_{0.05}$NiO$_2$

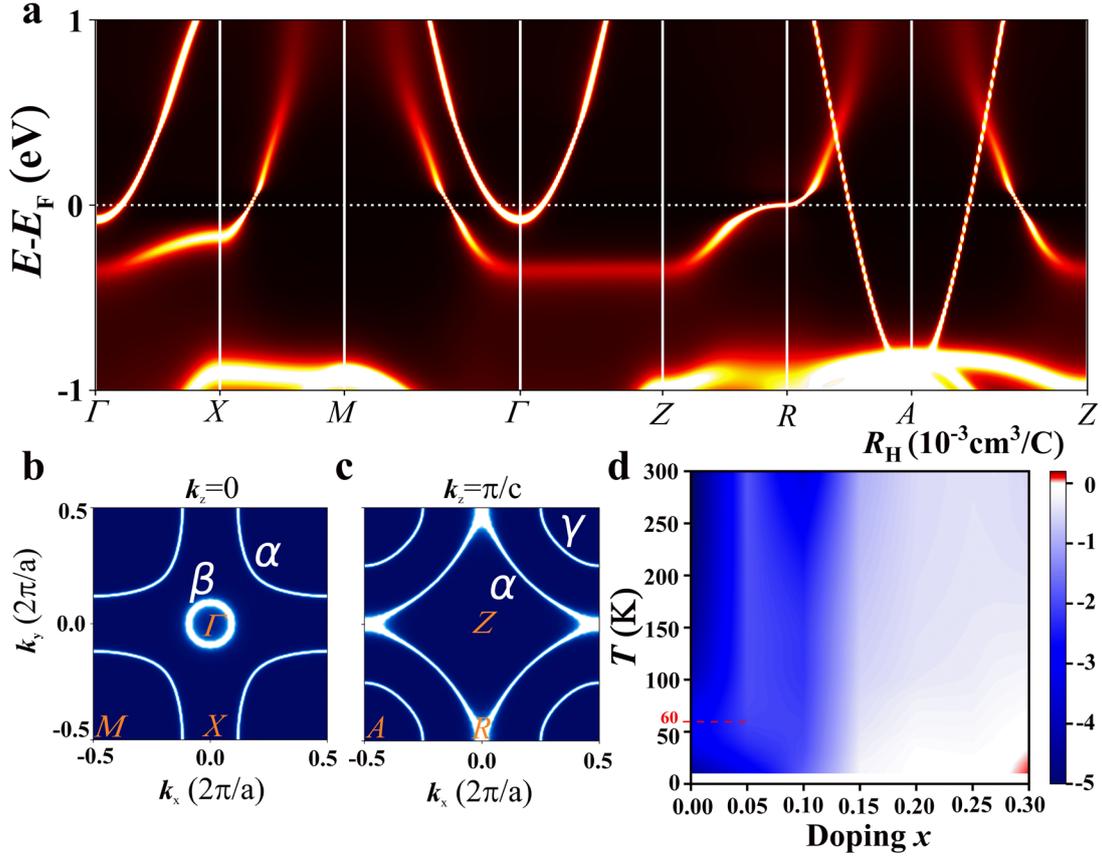

**Fig. 4 | Electronic structures of La$_{0.95}$Sr$_{0.05}$NiO$_2$ and Hall coefficients for all doping levels.** DFT+DMFT band structure (**a**) and two-dimensional Fermi surfaces in the $\Gamma$ plane (**b**) and $Z$ plane (**c**) of La$_{0.95}$Sr$_{0.05}$NiO$_2$. (**d**) The Contour map of experimental Hall coefficients ($R_H$) for various doping concentrations. The color gradient on the right side of the graph indicates the sign and magnitude of $R_H$. Blue color signifies negative values, white color represents zero, and red color denotes positive values.

VHS, as critical points in the electronic density of states due to the geometry of the electronic band structure, significantly influence electronic properties including carrier concentration. As evidenced by the doping-dependent evolution of the band structure shown in Figs. 1c-1e and 4a, the VHS at the $R$ point progressively shifts toward higher energies with increasing doping concentration. A Lifshitz transition is triggered when the VHS intersects the Fermi level at a critical doping level of $x=0.05$, concomitant with the reconfiguration of the $\alpha$ Fermi surface from enclosing the $A$ point to enclosing the $Z$ point in the $k_z =\pi/c$ plane (Fig. 4c). As illustrated in Fig. 4d, the La-based ILNs display a non-monotonic evolution of the $R_H$ magnitude across the doping range $x=0$ to $x=0.1$ at high temperatures ($T >60$ K). Strikingly, the magnitude of $|R_H|$ (absolute value) at $x=0.05$ exhibits a marked reduction compared to that at $x=0.1$ in high temperature range. This observation, when interpreted through the single-band carrier concentration $n = 1/(eR_H)$, where $n$ is the carrier concentration, $e$ is the elementary charge, suggests a higher carrier concentration at $x = 0.05$ than at $x = 0.1$. The correlation between $R_H$ anomalies and $n$ discontinuities represents a defining characteristic of VHS, as exemplified in graphene[68] and cuprates[69]. In ILNs, the doping level triggering this anomaly matches precisely with the theoretically predicted threshold of VHS-associated critical doping. This exact correspondence establishes compelling evidence for VHS formation in ILNs.

**Electronic susceptibilities in the $La_{1-x}Sr_xNiO_2$**

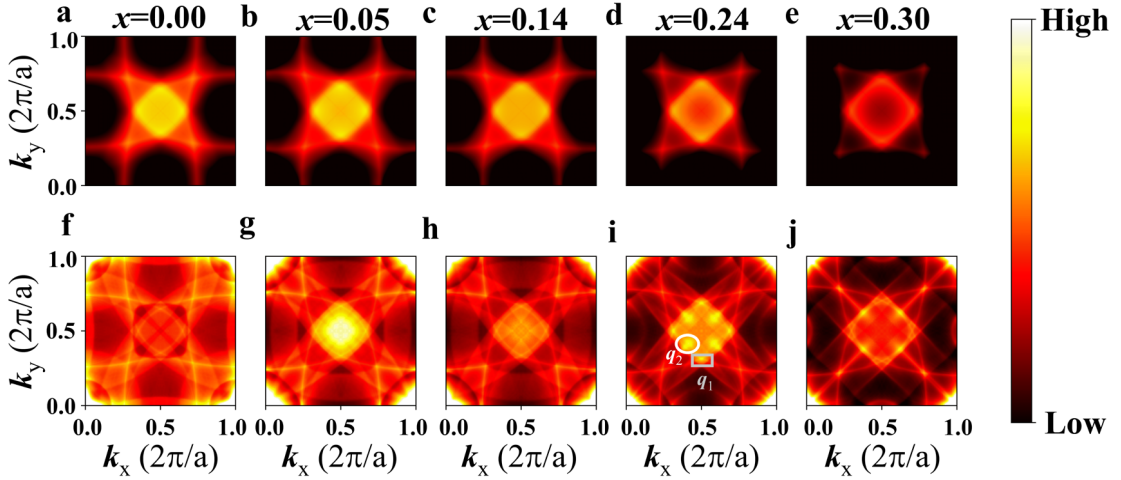

**Fig. 5 | Electronic susceptibilities in ILNs.** The real part (top row) and imagine part (bottom row) of electronic susceptibilities of ILNs in the $q_z$=0 plane with different hole doping concentrations $x$=0, 0.05, 0.14, 0.24, 0.3 (from left to right, respectively).

Figs. 5a-5j present the calculated electronic susceptibility, which is commonly employed to investigate the formation of CDWs[46,70,71]. The real and imaginary components of electronic susceptibility correspond to the static and dynamic instabilities of the electron system, respectively. FSN is identified as a credible origin for the emergence of CDW, with the phenomenon being delineated by the identification of extremum in the calculated electronic susceptibility within the reciprocal lattice space ($q$-space), thereby elucidating the FSN characteristics[72]. The electronic susceptibility is defined by its real and imaginary parts as follows:

$$Re[\chi(q)] = \sum_k \frac{f(\varepsilon_{k+q}) - f(\varepsilon_k)}{\varepsilon_k - \varepsilon_{k+q}} \quad (1)$$

where $\varepsilon_k$ is the eigenvalue of the electronic energy with a wave vector $k$, $f(\varepsilon_k)$ is the Fermi-Dirac distribution function. The FSN function is estimated by the imaginary part of the electronic susceptibility, and can be defined as:

$$Im[\chi(q)] = \sum_k \delta(\varepsilon_{k+q} - \varepsilon_F)\delta(\varepsilon_k - \varepsilon_F) \quad (2)$$

where $\varepsilon_F$ is the Fermi energy. When both the real and imaginary parts reach a peak at a wave vector $q$, it is considered to have the potential to form a CDW order at some lower temperature with $q_{CDW}=q$.

Given that the ILNs under investigation are ultra-thin films, our primary focus is on analyzing the electronic susceptibility in the $q_z$=0 plane. From Figs. 5a-5e, the calculated results show that near the $M$ point in the Brillouin zone, $Re[\chi(q)]$ demonstrates a prominent peak intensity. Moreover, this intensity marginally decreases as the doping concentration rises. This phenomenon indicates that the electronic susceptibility in the vicinity of the *M* point is highly sensitive to charge density modulation. When $x=0.24$, both $Re[\chi(q)]$ and $Im[\chi(q)]$ are notably maximized at $q_1$(0.5 $\frac{2\pi}{a}$, 0.31 $\frac{2\pi}{a}$, 0) and $q_2$(0.41 $\frac{2\pi}{a}$, 0.41 $\frac{2\pi}{a}$, 0). These peaks are not as strong as the FSN observed in other CDW systems, where the intensity at $q_{CDW}$ is comparable to that at the *Γ* point. While these peak of $Im[\chi(q)]$ are not strong enough to drive CDW order, they may be sufficient to cause the RSB in $La_{1-x}Sr_xNiO_2$. The unequal *x* and *y* components of $q_1$ implies that the scattering between the Fermi surfaces become unequal along the *x* and *y* directions, thereby violating $C_4$ symmetry. In addition, when $x=0.05$, although the intensity of $Im[\chi(q)]$ at $q_1$ point is not as large as that near *M* point, the maximum value of $Re[\chi(q)]$ is still at $q_1$ point. Therefore, we believe that the large $Im[\chi(q_1)]$ causes RSB at $x=0.05$. The corresponding nesting pattern of $q_1$ mainly comes from the mutual nesting of the rhombic *α* pockets near the $k_z=\pi/c$ plane and the *γ* pocket that is tangential to the *α* pocket. (The analysis of the nesting pattern provided in the Supplementary Information V.) From the comparison of Figs. 5f, 5h, 5i and 5j, it is easy to observe that at $x=0.24$, $Im[\chi(q)]$ exhibits sharp peaks at $q_1$ and $q_2$, whereas it is much weaker at $x=0$, 0.14 and 0.3.

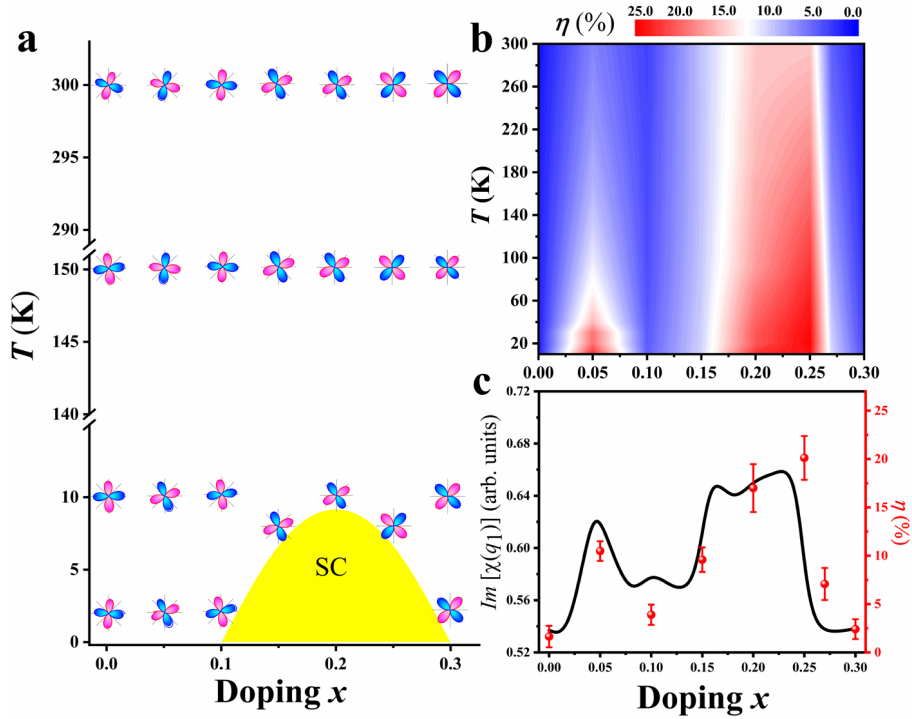

**Fig. 6 | Relationship between electronic susceptibilities and experimental RSB strength *η*.** (a) The temperature and doping dependence of $\rho_T(\varphi)$. The petal-shaped $\rho_T(\varphi)$ dependence is ubiquitous throughout the entire temperature and doping range. The radius is normalized to the same size. The yellow area represents the superconducting region. (b) The Contour map of RSB strength (anisotropy ratio *η* of resistivity) as a function of doping and temperature, with doping concentration *x* ranging from 0-0.3 and temperature ranging from 10-300 K. (c) The black curve illustrates the value of the imaginary part of the electronic susceptibility around $q_1$=(0.5 $\frac{2\pi}{a}$, 0.31 $\frac{2\pi}{a}$, 0) at 116 K, while the red points

correspond to experimentally measured RSB strength $\eta$ at 120 K. All experimental data points are annotated with vertical error bars. The detailed calculation of error bar is shown in Supplementary Information VI.

In Fig. 6a, we observed the doping dependence of $\rho_T(\varphi)$ petal shape from the parent compound to the overdoped region across various temperature ranges. From the phase diagram, it is also easy for us to observe that $\theta$ (see Fig.3c for definition) changes with both doping and temperature, and it is usually not aligned with the crystal orientations of [100], [010], [110], *etc.*, in the lattice, nor is fixed at a specific angle. Fig. 6b shows the doping and temperature dependent phase diagram of the RSB strength $\eta$. The phase diagram reveals that the RSB strength reaches its peak around $x = 0.05$ and $0.25$. In contrast, in cuprate superconductors, the RSB strength is most pronounced in the underdoped region[4]. Fig. 6c reveals a general alignment of the experimental RSB strength and the calculated $Im[\chi(q_1)]$ over the whole doping range. Comparative evaluation of $Im[\chi(q_1)]$ peak intensities at $q_1$ across different doping levels in ILNs reveals two prominent FSN plateaus: a relatively small peak near $x=0.05$ and a pronounced maximum at $x=0.24$. Both regimes exhibit significantly enhanced nesting strength compared to adjacent doping concentrations. Detailed nesting pattern analysis (Supplementary Information V) uncovers distinct physical origins for the enhancements at $x=0.05$ and $x=0.24$: The nesting amplification at $x=0.05$ derives from the migration of VHS at $R$-point toward the Fermi level, which geometrically optimizes the nesting conditions between equivalent $R$ points with $q=(0.5\frac{2\pi}{a}, 0.5\frac{2\pi}{a}, 0)$. This change of the electronic topology directly elevates $Im[\chi(q)]$ magnitudes in the vicinity of $M$ point, including $q_1$. The pronounced peak observed at $x=0.24$ originates from two factors: (1) In the $k_z=\pi/c$ plane, the $\alpha$ Fermi surface undergoes a topological transformation: from a circular Fermi surface surrounding $A$ point to a quasi-square Fermi surface surrounding $Z$ point. This geometric configuration facilitates parallel alignment of multiple Fermi surface sections, significantly improving nesting vector matching across Brillouin zone boundaries. (2) The $\gamma$ pocket develops tangential contact with the convex segments of the $\alpha$ Fermi surface, creating additional nesting channels. This inter-band interaction introduces supplementary momentum transfer pathways. The synergistic superposition of these intra-band ($\alpha$-$\alpha$) and inter-band ($\alpha$-$\gamma$) nesting configurations collectively contribute to the maximal nesting strength observed near the 0.24 doping level. The remarkable correspondence between the theoretically predicted peak of nesting strength ($x=0.05, 0.24$) and the experimentally determined doping level ($x=0.05, 0.25$), strongly suggests that electronic instability induced by FSN constitutes a primary driving mechanism for RSB in $La_{1-x}Sr_xNiO_2$.

## Conclusion

Spontaneous RSB is a typical characteristic of unconventional superconductors, manifested by the fact that the symmetry of the order parameter is lower than that of the lattice. We conducted the ARR and Hall effect experiments to meticulously investigate the anisotropy behavior in the normal state from the parent to the overdoped region. The experimental results indicate that in the normal state of $La_{1-x}Sr_xNiO_2$, there are significant peaks in the RSB strength at $x=0.05$ and $0.25$. Notably, this result differs from the phase diagram of the variation of RSB strength with doping in the normal state of cuprate films[4]. Based on the above experimental findings, we calculated the electronic susceptibilities of ILNs at different doping concentrations. A significant peak in the electronic susceptibility often indicates the presence of electronic instability in the system, and electrons tend to form periodic density modulations. The calculations show that the imaginary part of the electronic susceptibility exhibits a significant enhancement at doping concentrations of $x=0.05$ and $0.24$. The theoretical calculation results are in good agreement with the maximum RSB strength observed experimentally at the corresponding doping levels. Specifically, we identified a significant peak in the imaginary part of the electronic susceptibility at the wave vector $q_1(0.5\frac{2\pi}{a}, 0.31\frac{2\pi}{a}, 0)$, which is a typical characteristic of FSN. Particularly importantly, the $x$ and $y$ components of the $q_1$ peak are not equal, which directly leads to the breaking of $C_4$ symmetry and reasonably explains the RSB observed in the experiment. Through in-depth analysis of the nesting patterns, we found that the causes of anisotropy at $x=0.05$ and $0.24$ are different. At $x=0.05$, the anomalous increase in carrier concentration above 60K suggests the possible formation of VHS. This VHS induces the dominating FSN between the α bands at equivalent $R$ points, and leads to electronic instability. At $x=0.24$, the observed anisotropy is mainly affected by the FSN within the α−α band and between the α−γ bands. These findings provide a solid basis for in-depth understanding of RSB in ILNs.

## Methods

**Thin-film synthesis.** Perovskite precursor phase thin films were grown by PLD (KrF, λ = 248 nm, 4 Hz) onto $SrTiO_3(001)$ substrates, using stoichiometric polycrystalline targets[27,33,34]. During growth, the substrate temperature was kept at 520 °C. The oxygen pressure and laser fluence during the growth was 300 mTorr, 2.0 J/cm² and 150 mTorr, 1.0 J/cm² for undoped $LaNiO_3$ and doped $La_{1-x}Sr_xNiO_3$ ($x \neq 0$), respectively. The $La_{1-x}Sr_xNiO_3$ thin films were wrapped in aluminum foil and sealed in a quartz tube along with 0.5g $CaH_2$ powder. The glass tube was evacuated to 0.015 mTorr and then sealed. Subsequently, the glass tube was heated in a tube furnace at 320 °C for 4 hours to ensure that $La_{1-x}Sr_xNiO_3$ thin films were fully reduced single crystal ILNs. The heating and cooling rates were both 10 °C/min. The metal Nb film were grown on $SrTiO_3(001)$ substrates by RF magnetron sputter deposition with a sputtering power of 150 W at room temperature. We prepared a vacuum of a base pressure of $7.5 \times 10^{-5}$ mTorr and deposited films at an Ar pressure of 3.5 mTorr.

**Electrical transport measurements and lithography.** The electrical transport measurement is conducted on the commercial instrument Physical Property Measurement System (PPMS). Two Keithley 2400 DC current sources are utilized to apply rotating current density vectors, while two Keithley 2182A nano-voltmeters measured the longitudinal and the transverse voltages, respectively. To prevent self-heating of the device, we maintaine the excitation current density for the transport measurement at a sufficiently low level of 0.06366 A·cm$^{-2}$, which corresponds to a total current of 20 μA.

The ARR lithography pattern, first proposed by Chen et al.[5], is an innovation based on the "sun beam shaped" lithography pattern developed by Wu and his colleagues[4,73-75]. The photolithography pattern has a stripe width of 75 μm, a central circular diameter of 200 μm, and a square electrode length of 300 μm for wire bonding. This pattern allows for the rotation of the current density vector $J$ from 0° to 360° by controlling the current in the $x$ and $y$ directions, without the need for multiple wiring, achieving a resolution of up to 0.1°. The devices shown in the Extended Data Fig.1 are formed on La$_{1-x}$Sr$_x$NiO$_2$ thin film by using laser direct writing lithography technology. 5*3 mm$^2$ single-sided polished SrTiO$_3$(001) substrate, with its long edge oriented along the [100] direction, achieving an accuracy superior to ±1°. We align the lithography pattern with the edge of the substrate under a microscope, and the deviation between the $\theta$=0° direction and the crystallographic [100] direction is less than ±1°. Then wet etching is used here instead of ion-beam etching to prevent highly conductive nanolayers on SrTiO$_3$[76]. For ILNs, a 5.5% nitric acid aqueous solution was used to etch the thin film[77]. For Nb metal, we use the buffered chemical polishing method with a mixture of hydrofluoric acid (49%), nitric acid (70%), and phosphoric acid (85%) in a volume ratio of 1:1:1 to etch the film[78]. Finally, we remove the remaining photoresist with acetone. The transverse and longitudinal $J_x$, $J_y$ on the cross-section of the device is independently controlled by the current injected into the $I_{x+}$, $I_{x-}$ and $I_{y+}$, $I_{y-}$ contacts, respectively. $\varphi$ represents the angle between $J$ and [100] directions of the SrTiO$_3$ substrate. By setting $I_x=I_0\cos\varphi$ and $I_y=I_0\sin\varphi$, $J$ can be rotated within the plane at any angle interval $\varphi$, achieving a precision of 0.1° covering the complete range from $\varphi$=0° to 360°. The longitudinal voltage in the $J$ direction is $V_L = V_x\cos\varphi + V_y\sin\varphi$, while the transverse voltage in the $J$ direction is $V_T = -V_x\sin\varphi+V_y\cos\varphi$.

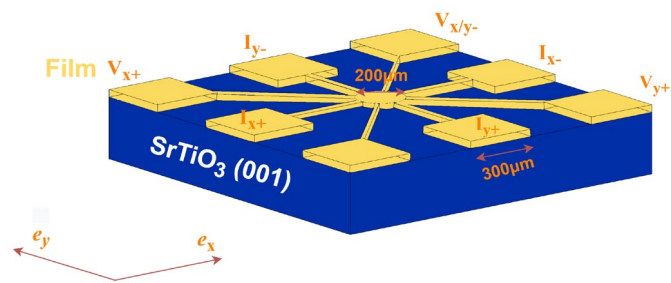

**Extended Data Fig. 1 | The lithography pattern for the ARR method.**

**Transverse voltage induced by anisotropic resistivity and the magnitude of anisotropy.** The differential forms of Ohm's law, $\mathbf{E}=\rho\mathbf{J}$, where $\rho$ is a tensor. For materials with orthogonal properties, the in-plane $\rho$ can be expressed as

$$\rho = \begin{bmatrix} \rho_a & 0 \\ 0 & \rho_b \end{bmatrix}$$

Here $\mathbf{e}_a$ and $\mathbf{e}_b$ are the principal axes. Let us define $\bar{\rho}= (\rho_a + \rho_b)/2$ and $\Delta\rho= (\rho_a - \rho_b)/2$; If the axis rotates by an angle $\varphi$, resulting in $\mathbf{e}_x=\hat{\mathbf{C}}_\varphi \mathbf{e}_a$ and $\mathbf{e}_y=\hat{\mathbf{C}}_\varphi \mathbf{e}_b$, in the $\mathbf{e}_x$ and $\mathbf{e}_y$ coordinate system, the resistivity tensor becomes

$$\hat{\mathbf{C}}_\varphi \rho \hat{\mathbf{C}}_\varphi^{-1} = \begin{pmatrix} \cos\varphi & -\sin\varphi \\ \sin\varphi & \cos\varphi \end{pmatrix} \begin{pmatrix} \rho_a & 0 \\ 0 & \rho_b \end{pmatrix} \begin{pmatrix} \cos\varphi & \sin\varphi \\ -\sin\varphi & \cos\varphi \end{pmatrix} = \begin{pmatrix} \bar{\rho}+\Delta\rho\cos(2\varphi) & \Delta\rho\sin(2\varphi) \\ \Delta\rho\sin(2\varphi) & \bar{\rho}-\Delta\rho\cos(2\varphi) \end{pmatrix}$$

In general, the principal axes $\mathbf{e}_a$ and $\mathbf{e}_b$ are frequently misaligned with the crystallographic [100] and [010] directions. The angle $\theta$, representing the deviation of the principal axis $\mathbf{e}_a$ from the [100] orientation of the SrTiO$_3$ substrate, is defined as the parameter specifying the anisotropic direction (see Fig. 3c). Therefore, the angle dependence of anisotropic electrical transport $\rho(\varphi)$ and $\rho_T(\varphi)$ should adhere to the following relationship[4,5]: $\rho_T(\varphi) = \Delta\rho\sin[2(\varphi-\theta)]$, $\rho(\varphi) = \bar{\rho} + \Delta\rho\cos[2(\varphi-\theta)]$. The RSB strength $\eta$ is defined in the following form: $\eta=[(\rho_a - \rho_b)/2]/[(\rho_a + \rho_b)/2] =\Delta\rho/\bar{\rho}$. By analyzing the vertical offset, amplitude and phase offset of angular oscillation, as well as $\bar{\rho}$, $\Delta\rho$ and $\theta$, we investigate the anisotropy in ILNs.

**Electronic structure calculation.** We use the DFT+DMFT[79] as implemented in the work of Haule et al.[80], which is based on the full potential linear augmented plane wave method implemented in Wien2K[81], to carry out our first principles calculations. A 20×20×23 $k$-point grid is used for self-consistent calculations. The quantum impurity problem is solved by the continuous time quantum Monte Carlo method[82,83]. All the five Ni-3$d$ orbitals are considered as correlated ones. The on-site Coulomb repulsion $U$ and the Hund's coupling $J_H$ are fixed at 5 eV and 1 eV, respectively, which is consistent with previous DFT+DMFT calculations of the ILN superconductors[84,85]. To mimic the experimental conditions for the deposition of La$_{1-x}$Sr$_x$NiO$_2$ thin films on SrTiO$_3$ substrates, the in-plane lattice constants ($a$ and $b$ axes) of La$_{1-x}$Sr$_x$NiO$_2$ are fixed to match those of the SrTiO$_3$ substrate (3.91 Å)[20]. The out-of-plane $c$-axis lattice constant is set to 3.422 Å, as determined by structural relaxation calculation. The doping effects are investigated using the virtual crystal approximation method.

### Data availability
All data necessary for assessing the conclusions drawn in this study are included within the manuscript and the Supplementary Information. Data underpinning the findings of this research can be obtained from the corresponding author upon formal request.

### References

1   Kivelson, S. A., Fradkin, E. & Emery, V. J. Electronic liquid-crystal phases of a doped Mott insulator. *Nature* **393**, 550-553, (1998).

2	Oganesyan, V., Kivelson, S. A. & Fradkin, E. Quantum theory of a nematic Fermi fluid. *Phys. Rev. B* **64**, 6, (2001).

3	Hinkov, V. *et al.* Electronic liquid crystal state in the high-temperature superconductor YBa$_2$Cu$_3$O$_{6.45}$. *Science* **319**, 597-600, (2008).

4	Wu, J., Bollinger, A. T., He, X. & Bozovic, I. Spontaneous breaking of rotational symmetry in copper oxide superconductors. *Nature* **547**, 432-435, (2017).

5	Chen, G. F. *et al.* Angle-resolved resistivity method for precise measurements of electronic nematicity. *Phys. Rev. B* **106**, 6, (2022).

6	Abdel-Jawad, M. *et al.* Anisotropic scattering and anomalous normal-state transport in a high-temperature superconductor. *Nat. Phys.* **2**, 821-825, (2006).

7	Li, J. *et al.* Nematic superconducting state in iron pnictide superconductors. *Nat. Commun.* **8**, 8, (2017).

8	Chu, J. H. *et al.* In-Plane Resistivity Anisotropy in an Underdoped Iron Arsenide Superconductor. *Science* **329**, 824-826, (2010).

9	Wang, J. H. *et al.* Progress of nematic superconductivity in iron-based superconductors. *Adv. Phys.-X* **6**, 16, (2021).

10	Xiang, Y. *et al.* Twofold symmetry of *c*-axis resistivity in topological kagome superconductor CsV$_3$Sb$_5$ with in-plane rotating magnetic field. *Nat. Commun.* **12**, 8, (2021).

11	Jiang, Z. C. *et al.* Flat bands, non-trivial band topology and rotation symmetry breaking in layered kagome-lattice RbTi$_3$Bi$_5$. *Nat. Commun.* **14**, 8, (2023).

12	Cao, Y. *et al.* Nematicity and competing orders in superconducting magic-angle graphene. *Science* **372**, 264-271, (2021).

13	Zhang, N. J. *et al.* Angle-resolved transport non-reciprocity and spontaneous symmetry breaking in twisted trilayer graphene. *Nat. Mater.* **23**, 316-322, (2024).

14	Zhang, G. Q. *et al.* Spontaneous rotational symmetry breaking in KTaO$_3$ heterointerface superconductors. *Nat. Commun.* **14**, (2023).

15	Davis, S. *et al.* Signatures of electronic nematicity in (111) LaAlO$_3$/SrTiO$_3$ interfaces. *Phys. Rev. B* **97**, 6, (2018).

16	Li, H. *et al.* Unidirectional coherent quasiparticles in the high-temperature rotational symmetry broken phase of *A*V$_3$Sb$_5$ kagome superconductors. *Nat. Phys.*, 637-643, (2023).

17	Jin, C. H. *et al.* Stripe phases in WSe$_2$/WS$_2$ moire superlattices. *Nat. Mater.* **20**, 940-944, (2021).

18	Chu, J. H., Kuo, H. H., Analytis, J. G. & Fisher, I. R. Divergent Nematic Susceptibility in an Iron Arsenide Superconductor. *Science* **337**, 710-712, (2012).

19	Ji, H. R. *et al.* Rotational symmetry breaking in superconducting nickelate Nd$_{0.8}$Sr$_{0.2}$NiO$_2$ films. *Nat. Commun.* **14**, 8, (2023).

20	Li, D. F. *et al.* Superconductivity in an infinite-layer nickelate. *Nature* **572**, 624-627, (2019).

21	Zeng, S. W. *et al.* Phase Diagram and Superconducting Dome of Infinite-Layer Nd$_{1-x}$Sr$_x$NiO$_2$ Thin Films. *Phys. Rev. Lett.* **125**, 7, (2020).

22   Lee, K. *et al.* Linear-in-temperature resistivity for optimally superconducting (Nd,Sr)NiO$_2$. *Nature* **619**, 288-292, (2023).

23   Rossi, M. *et al.* A broken translational symmetry state in an infinite-layer nickelate. *Nat. Phys.* **18**, 869-873, (2022).

24   Parzyck, C. T. *et al.* Absence of 3a$_0$ charge density wave order in the infinite-layer nickelate NdNiO$_2$. *Nat. Mater.* **23**, 8, (2024).

25   Yan, S. J. *et al.* Superconductivity in Freestanding Infinite-Layer Nickelate Membranes. *Adv. Mater.* **36**, 8, (2024).

26   Wei, W. *et al.* Large upper critical fields and dimensionality crossover of superconductivity in the infinite-layer nickelate La$_{0.8}$Sr$_{0.2}$NiO$_2$. *Phys. Rev. B* **107**, 7, (2023).

27   Zhao, Q. *et al.* Isotropic Quantum Griffiths Singularity in Nd$_{0.8}$Sr$_{0.2}$NiO$_2$ Infinite-Layer Superconducting Thin Films. *Phys. Rev. Lett.* **133**, 6, (2024).

28   Ding, X. *et al.* Cuprate-like electronic structures in infinite-layer nickelates with substantial hole dopings. *Natl. Sci. Rev.* **11**, 7, (2024).

29   Lu, H. *et al.* Magnetic excitations in infinite-layer nickelates. *Science* **373**, 213-216, (2021).

30   Yang, Z. *et al.* Photoinduced Phase Transition in Infinite-Layer Nickelates. *Small* **19**, 7, (2023).

31   Zeng, S. W. *et al.* Observation of perfect diamagnetism and interfacial effect on the electronic structures in infinite layer Nd$_{0.8}$Sr$_{0.2}$NiO$_2$ superconductors. *Nat. Commun.* **13**, 6, (2022).

32   Zhang, G. M., Yang, Y. F. & Zhang, F. C. Self-doped Mott insulator for parent compounds of nickelate superconductors. *Phys. Rev. B* **101**, 5, (2020).

33   Nie, J. *et al.* Irrelevance of $^1$H composition to superconductivity in infinite-layer nickelates from measurements of nuclear interactions. *Newton* **1**, 100006, (2025).

34   Shao, T. N. *et al.* Kondo scattering in underdoped Nd$_{1-x}$Sr$_x$NiO$_2$ infinite-layer superconducting thin films. *Natl. Sci. Rev.* **10**, 7, (2023).

35   Zhao, D. *et al.* Intrinsic Spin Susceptibility and Pseudogaplike Behavior in Infinite-Layer LaNiO$_2$. *Phys. Rev. Lett.* **126**, 8, (2021).

36   Lin, H. *et al.* Universal spin-glass behaviour in bulk LaNiO$_2$, PrNiO$_2$ and NdNiO$_2$. *New J. Phys.* **24**, 11, (2022).

37   Ortiz, R. A. *et al.* Magnetic correlations in infinite-layer nickelates: An experimental and theoretical multimethod study. *Phys. Rev. Res.* **4**, 19, (2022).

38   Fowlie, J. *et al.* Intrinsic magnetism in superconducting infinite-layer nickelates. *Nat. Phys.* **18**, 1043-1047, (2022).

39   Chen, Z. Y. *et al.* Electronic structure of superconducting nickelates probed by resonant photoemission spectroscopy. *Matter* **5**, 11, (2022).

40   Sakakibara, H. *et al.* Model Construction and a Possibility of Cupratelike Pairing in a New $d^9$ Nickelate Superconductor (Nd,Sr)NiO$_2$. *Phys. Rev. Lett.* **125**, 6, (2020).

41   Botana, A. S. & Norman, M. R. Similarities and Differences between LaNiO$_2$ and CaCuO$_2$ and Implications for Superconductivity. *Phys. Rev. X* **10**, 6, (2020).

42  Tam, C. C. *et al.* Charge density waves in infinite-layer $NdNiO_2$ nickelates. *Nat. Mater.* **21**, 1116-1120, (2022).

43  Pelliciari, J. *et al.* Comment on newly found Charge Density Waves in infinite layer Nickelates. arXiv:2306.15086 (2023).

44  Tam, C. C. *et al.* Reply to "Comment on newly found Charge Density Waves in infinite layer Nickelates". arXiv:2307.13569 (2023).

45  Rossi, M. *et al.* Universal orbital and magnetic structures in infinite-layer nickelates. *Phys. Rev. B* **109**, 9, (2024).

46  Sui, X. L. *et al.* Hole doping dependent electronic instability and electron-phonon coupling in infinite-layer nickelates. *Phys. Rev. B* **107**, 19, (2023).

47  Raji, A. *et al.* Charge Distribution across Capped and Uncapped Infinite-Layer Neodymium Nickelate Thin Films. *Small* **19**, 7, (2023).

48  Ando, Y., Segawa, K., Komiya, S. & Lavrov, A. N. Electrical resistivity Anisotropy from self-organized one dimensionality in high-temperature superconductors. *Phys. Rev. Lett.* **88**, 4, (2002).

49  Borzi, R. A. *et al.* Formation of a nematic fluid at high fields in $Sr_3Ru_2O_7$. *Science* **315**, 214-217, (2007).

50  Ronning, F. *et al.* Electronic in-plane symmetry breaking at field-tuned quantum criticality in $CeRhIn_5$. *Nature* **548**, 313-317, (2017).

51  Li, L., Alidoust, N., Tranquada, J. M., Gu, G. D. & Ong, N. P. Unusual Nernst Effect Suggesting Time-Reversal Violation in the Striped Cuprate Superconductor $La_{2-x}Ba_xCuO_4$. *Phys. Rev. Lett.* **107**, 5, (2011).

52  Nie, L. P. *et al.* Charge-density-wave-driven electronic nematicity in a kagome superconductor. *Nature* **604**, 59-64, (2022).

53  Walmsley, P. & Fisher, I. R. Determination of the resistivity anisotropy of orthorhombic materials via transverse resistivity measurements. *Rev. Sci. Instrum.* **88**, 9, (2017).

54  Kivelson, S. A. *et al.* How to detect fluctuating stripes in the high-temperature superconductors. *Rev. Mod. Phys.* **75**, 1201-1241, (2003).

55  Fradkin, E., Kivelson, S. A. & Tranquada, J. M. *Colloquium*: Theory of intertwined orders in high temperature superconductors. *Rev. Mod. Phys.* **87**, 457-482, (2015).

56  Dagotto, E. Complexity in strongly correlated electronic systems. *Science* **309**, 257-262, (2005).

57  Emery, V. J., Kivelson, S. A. & Tranquada, J. M. Stripe phases in high-temperature superconductors. *Proc. Natl. Acad. Sci. U. S. A.* **96**, 8814-8817, (1999).

58  Hanaguri, T. *et al.* A 'checkerboard' electronic crystal state in lightly hole-doped $Ca_{2-x}Na_xCuO_2Cl_2$. *Nature* **430**, 1001-1005, (2004).

59  Kohsaka, Y. *et al.* An intrinsic bond-centered electronic glass with unidirectional domains in underdoped cuprates. *Science* **315**, 1380-1385, (2007).

60  Pan, S. H. *et al.* Microscopic electronic inhomogeneity in the high-$T_c$ superconductor $Bi_2Sr_2CaCu_2O_{8+x}$. *Nature* **413**, 282-285, (2001).

61  Jiang, H. C. & Kivelson, S. A. Stripe order enhanced superconductivity in the

| | |
|---|---|
| | Hubbard model. *Proc. Natl. Acad. Sci. U. S. A.* **119**, 6, (2022). |
| 62 | Lane, C. *et al.* Competing incommensurate spin fluctuations and magnetic excitations in infinite-layer nickelate superconductors. *Commun. Phys.* **6**, 11, (2023). |
| 63 | Zhang, R. Q. *et al.* Emergence of Competing Stripe Phases in Undoped Infinite-Layer Nickelates. *Phys. Rev. Lett.* **133**, 7, (2024). |
| 64 | Sun, W. J. *et al.* Electronic structure of superconducting infinite-layer lanthanum nickelates. *Sci. Adv.* **11**, 8, (2025). |
| 65 | Choi, M. Y., Pickett, W. E. & Lee, K. W. Fluctuation-frustrated flat band instabilities in $NdNiO_2$. *Phys. Rev. Res.* **2**, 7, (2020). |
| 66 | Leonov, I., Skornyakov, S. L. & Savrasov, S. Y. Lifshitz transition and frustration of magnetic moments in infinite-layer $NdNiO_2$ upon hole doping. *Phys. Rev. B* **101**, 5, (2020). |
| 67 | Xu, M. Y. *et al.* Superconductivity Favored Anisotropic Phase Stiffness in Infinite-Layer Nickelates. arXiv:2502.14633 (2025). |
| 68 | Ma, C. X. *et al.* Evidence of van Hove Singularities in Ordered Grain Boundaries of Graphene. *Phys. Rev. Lett.* **112**, 5, (2014). |
| 69 | Maharaj, A. V., Esterlis, I., Zhang, Y., Ramshaw, B. J. & Kivelson, S. A. Hall number across a van Hove singularity. *Phys. Rev. B* **96**, 16, (2017). |
| 70 | Zhu, X. T., Cao, Y. W., Zhang, J. D., Plummer, E. W. & Guo, J. D. Classification of charge density waves based on their nature. *Proc. Natl. Acad. Sci. U. S. A.* **112**, 2367-2371, (2015). |
| 71 | Laverock, J. *et al.* Fermi surface nesting and charge-density wave formation in rare-earth tritellurides. *Phys. Rev. B* **71**, 5, (2005). |
| 72 | Johannes, M. D. & Mazin, II. Fermi surface nesting and the origin of charge density waves in metals. *Phys. Rev. B* **77**, 8, (2008). |
| 73 | Wu, J. *et al.* Electronic nematicity in $Sr_2RuO_4$. *Proc. Natl. Acad. Sci. U. S. A.* **117**, 10654-10659, (2020). |
| 74 | Wu, J., Bollinger, A. T., He, X. & Bozovic, I. Detecting Electronic Nematicity by the Angle-Resolved Transverse Resistivity Measurements. *J. Supercond. Nov. Magn* **32**, 1623-1628, (2019). |
| 75 | Wu, J. *et al.* Angle-Resolved Transport Measurements Reveal Electronic Nematicity in Cuprate Superconductors. *J. Supercond. Nov. Magn* **33**, 87-92, (2020). |
| 76. | Reagor, D. W. & Butko, V. Y. Highly conductive nanolayers on strontium titanate produced by preferential ion-beam etching. *Nat. Mater.* **4**, 593-596, (2005). |
| 77. | Osada, M. *et al.* A Superconducting Praseodymium Nickelate with Infinite Layer Structure. *Nano Lett.* **20**, 5735-5740, (2020). |
| 78. | Aspart, A. & Antoine, C. Z. Study of the chemical behavior of hydrofluoric, nitric and sulfuric acids mixtures applied to niobium polishing. *Appl. Surf. Sci.* **227**, 17-29, (2004). |
| 79. | Kotliar, G. *et al.* Electronic structure calculations with dynamical mean-field theory. *Rev. Mod. Phys.* **78**, 865-951, (2006). |

80. Haule, K., Yee, C. H. & Kim, K. Dynamical mean-field theory within the full-potential methods: Electronic structure of CeIrIn$_5$, CeCoIn$_5$, and CeRhIn$_5$. *Phys. Rev. B* **81**, 30, (2010).
81. Blaha, P. *et al.* WIEN2k: An APW+lo program for calculating the properties of solids. *J. Chem. Phys.* **152**, 30, (2020).
82. Haule, K. Quantum Monte Carlo impurity solver for cluster dynamical mean-field theory and electronic structure calculations with adjustable cluster base. *Phys. Rev. B* **75**, 12, (2007).
83. Werner, P., Comanac, A., de' Medici, L., Troyer, M. & Millis, A. J. Continuous-time solver for quantum impurity models. *Phys. Rev. Lett.* **97**, 4, (2006).
84. Kang, C. J. & Kotliar, G. Optical Properties of the Infinite-Layer La$_{1-x}$Sr$_x$NiO$_2$ and Hidden Hund's Physics. *Phys. Rev. Lett.* **126**, 7, (2021).
85. Wang, Y., Kang, C. J., Miao, H. & Kotliar, G. Hund's metal physics: From SrNiO$_2$ to LaNiO$_2$. *Phys. Rev. B* **102**, 8, (2020).

## Acknowledgements

**Funding:** This work was supported by the National Natural Science Foundation of China (92065110, 11974048, 12074334, 12074041, and 12374037), the Fundamental Research Funds for the Central Universities, and the Innovation Program for Quantum Science and Technology (Grant No. 2021ZD0302800), and the National Key Research and Development Program of China (No. 2021YFA0718900). The calculations were carried out on the high-performance computing cluster of Beijing Normal University in Zhuhai.

**Author contributions:** J.C.N conceived the project and supervised the experiments. Z.P.Y is responsible for all theoretical calculations. Q.Z. and W.L.Y. grew nickelate thin films and conducted characterization measurements using XRD and PPMS. R.L. conducted theoretical calculations. C.X., J.K.C., Q.Z., W.L.Y., J.K.L., and J.Y.M. conducted FIB and TEM characterization work. Q.Z. analyzed the experimental data with W.L.Y., X.Y.W., F.H.Z., C.X.C., M.L.Y.. Q.Z. and R.L. wrote the manuscript with help from R.F.D., C.M.X., C.X., X.Y.L., H.W.L and J.K.C with input from all authors.

**Competing interests:** The authors declare no competing interests.